\documentclass[a4paper,11pt]{article}
\usepackage{jheppub} 
\usepackage{booktabs}
\usepackage{subcaption}
\newcommand{\dd}{\mathrm{d}}
\newcommand{\e}{\mathrm{e}}

\newcommand{\p}{\partial}

\preprint{TTK-26-28}

\title{Integration-by-parts identities in the presence of measurement delta functions}

\author{Saimeng Zhou}
\affiliation{Institut f\"ur Theoretische Teilchenphysik und Kosmologie,
  RWTH Aachen University,\\D-52056 Aachen, Germany}
\emailAdd{saimeng.zhou@rwth-aachen.de}

\abstract{We derive a convenient form of integration-by-parts (IBP) identities for dimensionally regulated loop integrals that include a measurement constraint imposed by a delta distribution $\delta(\phi(\ell))$, where $\phi$ is a regular and generally non-linear function of the loop momentum.
Non-trivial IBP relations can be generated directly by vectors tangent to the hypersurface $\phi=0$, leaving the measurement delta as an overall factor and avoiding derivatives of distributions.
This construction provides a systematic route to differential distributions within reverse unitarity for non-linear measurement functions, in particular, for transverse-momentum distributions.
We implement the construction and validate it for transverse-momentum distributions in $t\bar tH$ production at leading order and $t\bar t$ production at next-to-leading order in QCD.
}


\begin{document}
\maketitle
\flushbottom

\section{Introduction}
\label{sec:intro}

There are several methods for computing scattering cross sections at low orders of perturbation theory in QCD. At next-to-leading and next-to-next-to-leading order, subtraction \cite{CataniSeymour1997,CataniDittmaierSeymourTrocsanyi2002,FrixioneKunsztSigner1996,CzakonHeymes2014} and slicing \cite{CataniGrazzini2007,BoughezalFockeLiuPetriello2015,GauntStahlhofenTackmannWalsh2015} methods allow one to obtain results differential in arbitrary infrared-safe observables at lepton-lepton, lepton-hadron, and hadron-hadron colliders. However, the dependence of the cross section on the observable is obtained only by numerical integration over the phase space.

Reverse unitarity \cite{AnastasiouMelnikov2002} provides a complementary approach by rewriting on-shell delta functions as cut propagators and thereby recasting phase-space integrals as cut loop integrals. This makes the methods developed for multi-loop integrals directly applicable to phase-space integration. In particular, integration-by-parts (IBP) identities \cite{ChetyrkinTkachov1981,Laporta2000} reduce the resulting integrals to a finite basis of master integrals whose kinematic dependence can be determined from differential equations \cite{Kotikov1991,Remiddi1997,GehrmannRemiddi2000,Henn2013}. This strategy underlies, for example, inclusive Higgs production at N$^3$LO
\cite{AnastasiouDuhrDulatHerzogMistlberger2015,Mistlberger2018} and has also been applied to differential cross-section calculations, such as rapidity distributions \cite{AnastasiouDixonMelnikovPetriello2004}. For total cross sections, the cut momenta are integrated over completely. For differential cross sections, the relevant information about the final-state kinematics must instead be retained through an additional measurement constraint.

Such a constraint takes the form of a delta distribution $\delta(\phi(\ell))$
fixing the observable in terms of the integration momenta. If the measurement function $\phi$ is linear in the relevant scalar products, as it is for rapidity distributions, the constraint can be included as an additional cut propagator in a standard Laporta family. By contrast, a transverse-momentum constraint is bilinear in the two beam projections of the momentum of the observed particle. Promoting it to a denominator would lead to a non-linear map between propagators and scalar products, so the constraint cannot be incorporated in the same way.

Non-linear measurement constraints have nevertheless been incorporated within reverse unitarity by other means. For energy-energy correlators, for example, the measurement has been promoted to a non-linear cut propagator and the IBP system supplemented by an additional relation that lowers the power of this propagator
\cite{DixonLuoShtabovenkoYangZhu2018,LuoShtabovenkoYangZhu2019,LeeLiXuZhang2026}.
A different route is taken in the Higgs-differential approach of Refs.~\cite{DulatLionettiMistlbergerPelloniSpecchia2017,DulatMistlbergerPelloni2018}.
There, the phase space is factorized into the momentum of the color-singlet final state and the accompanying QCD radiation. The momentum of the produced Higgs is kept fully exclusive and enters the remaining integrals as an external momentum, while the radiation is integrated inclusively by reverse unitarity. A single-differential spectrum such as $\dd\sigma/\dd p_T$ is obtained by integrating over the unmeasured rapidity afterwards. This construction is effective when the final state contains a color-singlet system whose momentum can be factorized from the QCD radiation. It does not, however, directly provide an IBP system for a general measurement constraint involving the integration momenta themselves.

In this work we instead keep the measurement delta function explicitly inside the phase-space integral and formulate the IBP problem in its presence. We show that derivatives of the measurement distribution can be eliminated by choosing IBP-generating vectors tangent to the hypersurface $\phi=0$.\footnote{Tangent vectors to a sphere have been previously used in the special case of angular integrals in Refs.~\cite{HaugWunder2024,HaugSmirnovWunder2025}; see Section~\ref{sec:tangentIBP}.} The measurement delta function then remains as an overall factor, and the resulting identities take the form of ordinary IBP relations that, once generated, can be solved by conventional Laporta elimination \cite{Laporta2000}. The same tangent vectors supply the derivative operators required for differential equations, so that master integrals defined at fixed transverse momentum can also be evolved in the observable. These equations determine the kinematic dependence exactly. In the applications presented here, we solve them numerically.

We implement this construction as a source-level extension of \textsc{Kira}
\cite{MaierhoferUsovitschUwer2018,KlappertLangeMaierhoferUsovitsch2021,LangeUsovitschWu2025} and demonstrate it for two transverse-momentum distributions: the Higgs spectrum in $pp\to t\bar tH$ at leading order and the single-inclusive top-quark spectrum in $pp\to t\bar t$ at next-to-leading order in QCD. Both are known independently, which is what makes them useful as tests of the method. The IBP identities and the associated differential operators are derived in Section~\ref{sec:IBP}. The implementation, boundary conditions, validation, and computational cost are presented in Section~\ref{sec:applications}, and we conclude in Section~\ref{sec:conclusions}. Appendix~\ref{app:boundarydetails} collects details on the boundary conditions, and Appendix~\ref{app:othermeasurements} extends the construction to further non-linear measurement constraints, including the transverse recoil of a multiparticle system, rapidity differences, and energy-energy correlators.

\section{IBP identities for measurement-constrained integrals}
\label{sec:IBP}
A standard method of computing phase-space integrals is reverse unitarity \cite{AnastasiouMelnikov2002}, which turns phase-space integrals into cut loop integrals by rewriting delta functions as cut propagators using Cutkosky's rule \cite{Cutkosky:1960sp},
\begin{equation}
\label{eq:cutkosky}
  \delta\!\left(x\right)
  \;=\;
  -\frac{1}{2\pi i}\left(\frac{1}{x+i 0}-\frac{1}{x-i 0}\right).
\end{equation}
This enables the direct use of multi-loop technology, e.g., IBP reduction to master integrals and their evaluation using differential equations. To obtain distributions in an observable, one must impose additional measurement constraints, typically represented by a delta distribution $\delta(\phi(\ell))$ that fixes the observable as a function of the loop momentum.\footnote{One may also encounter step functions $\Theta(\phi)$, for which a treatment within reverse unitarity has been presented in \cite{BaranowskiDeltoMelnikovWang2022,BaranowskiDeltoMelnikovPikelnerWang2024,BaranowskiDeltoMelnikovPikelnerWang2025}.}
How easily $\delta(\phi)$ can be incorporated in IBP reductions depends on the functional form of $\phi$.

For example, for rapidity distributions, the measurement constraint for a momentum $p$ with respect to lightlike beam momenta $p_1,p_2$ is given by:
\begin{equation}
  y \;=\; \frac12 \ln\!\left(\frac{p_2\!\cdot\!p}{p_1\!\cdot\!p}\right),
  \qquad
  \delta(y-y_0)\;\propto\;\delta\!\left(p_2\!\cdot\!p - \e^{2y_0}\,p_1\!\cdot\!p\right),
\end{equation}
which is linear in the scalar products $p_1\!\cdot\!p$ and $p_2\!\cdot\!p$ and can therefore be treated exactly like an additional cut linear propagator \cite{AnastasiouDixonMelnikovPetriello2004}.

For transverse momentum, the constraint is bilinear in two independent scalar products.
For an on-shell momentum $p$ of mass $m$, a fixed-$p_T$ measurement corresponds to
\begin{equation}
\label{eq:pTdeltaintro}
  \delta\!\left(|\vec p_\perp|^2 - p_T^2\right)
  \;=\;
  \delta\!\left(\frac{4}{s}(p_1\!\cdot\!p)(p_2\!\cdot\!p) - (p_T^2+m^2)\right),
\end{equation}
where $s=2\,p_1\!\cdot\! p_2$. If one attempts to promote the argument of this delta function to an additional propagator in an integral family, one introduces a quadratic relation among scalar products.
Inverting the propagator map to express scalar products in terms of inverse propagators then leads to square-root branches, so that not all scalar products can be linearly expressed by a chosen set of inverse propagators required for a closed IBP reduction. This is the obstruction that prevents a straightforward cut-propagator implementation of $p_T$-distributions in IBP setups. 

A similar issue has been addressed for the energy-energy correlator (EEC) in $e^+e^-$ annihilation, whose angular measurement in $\chi$ can be written as
\begin{equation}
  \delta\Big((1-\cos\chi)\,(p_i\!\cdot\!p)(p_j\!\cdot\!p) - p^2\, p_i\!\cdot\!p_j\Big),
\end{equation}
with $p$ the total incoming momentum. There, the delta function was promoted to a non-linear cut propagator, and the reduction was closed by supplementing the IBP system with an additional relation that lowers the power of the corresponding cut propagator by one \cite{DixonLuoShtabovenkoYangZhu2018,LuoShtabovenkoYangZhu2019,LeeLiXuZhang2026}.

\medskip
The IBP identities derived in Section~\ref{sec:tangentIBP} provide a complementary route in which the measurement function is not promoted to a cut propagator. Instead, the delta function is kept as an overall factor throughout, ensuring the integrals remain within the same integral family across the IBP system.

\subsection{From a differential cross section to measurement-constrained integrals}
\label{sec:example}

To illustrate how measurement-constrained integrals arise, consider the Higgs transverse-momentum distribution in $pp\to t\bar tH$ at LO, which is one of the applications of Section~\ref{sec:applications}. The hadronic cross section is the convolution of the partonic cross section $\dd\hat\sigma_{ij}/\dd p_T^2$ with parton distribution functions. For $i(p_1)+j(p_2)\to t(\ell_1)+H(\ell_2)+\bar t(\ell_3)$,
\begin{align}
\label{eq:exampleXS}
  \frac{\dd\hat\sigma_{ij}}{\dd p_T^2}
  =\frac{1}{2s}\int\prod_{a=1}^{3}\frac{\dd^d\ell_a}{(2\pi)^{d-1}}\,\delta_+(\ell_a^2-m_a^2)\;
  &(2\pi)^d\,\delta^{d}(p_1+p_2-\ell_1-\ell_2-\ell_3)
  \nonumber\\
  &\times\overline{|\mathcal M_{ij}|^2}\;\delta\big(|\vec \ell_{2,\perp}|^2-p_T^2\big),
\end{align}
where $\delta_+(\ell^2-m^2)=\theta(\ell^0)\,\delta(\ell^2-m^2)$ and $m_1=m_3=m_t$, $m_2=m_H$, while $\overline{|\mathcal M_{ij}|^2}$ denotes the sum and average over spins and colors of the squared matrix element. Integrating over $\ell_3$ and writing the measurement as in Eq.~\eqref{eq:pTdeltaintro} gives
\begin{align}
\label{eq:exampleXS2}
  \frac{\dd\hat\sigma_{ij}}{\dd p_T^2}
  =\frac{\pi}{s}\int\frac{\dd^d\ell_1}{(2\pi)^{d-1}}\,\frac{\dd^d\ell_2}{(2\pi)^{d-1}}\;
  &\delta_+(\ell_1^2-m_t^2)\,\delta_+\big((p_1+p_2-\ell_1-\ell_2)^2-m_t^2\big)\,\delta_+(\ell_2^2-m_H^2)
  \nonumber\\
  &\times\overline{|\mathcal M_{ij}|^2}\;\delta\big(\phi(\ell_2)\big),
\end{align}
with $\phi(\ell_2)=\frac{4}{s}(p_1\!\cdot\!\ell_2)(p_2\!\cdot\!\ell_2)-(p_T^2+m_H^2)$. The squared matrix element is a rational function of the scalar products of the momenta, with denominators given by propagators of the contributing diagrams. Reverse unitarity replaces the three on-shell delta functions by cut propagators through Eq.~\eqref{eq:cutkosky}. After mapping onto integral families, the partonic cross section becomes a linear combination of two-loop integrals
\begin{equation}
\label{eq:exampleform}
  I=\int\dd^d\ell_1\,\dd^d\ell_2\;f(\ell_1,\ell_2)\,\delta\big(\phi(\ell_2)\big),
\end{equation}
where $f$ is a product of propagators and irreducible scalar products. The same steps apply to any infrared-safe observable that fixes a function of the final-state momenta while all other degrees of freedom are integrated over. This includes energy-energy correlators and distributions of color-singlet particles or heavy quarks, such as those in Section~3, but not observables with kinematic cuts or jet clustering.

\subsection{IBP identities from tangent vectors}
\label{sec:tangentIBP}

We consider an integral constrained by a measurement function,
\begin{equation}
\label{eq:Imeas}
  I[f,\delta(\phi)] \;=\; \int \dd^d \ell \; f(\ell)\,\delta\big(\phi(\ell)\big),
\end{equation}
where $\phi(\ell)$ is a sufficiently regular scalar function, not assumed to be linear and not restricted to on-shell constraints, and $f(\ell)$ is a function built from scalar propagators for which the usual IBP construction applies. The measurement delta localizes the integration to the hypersurface
\begin{equation}
  \Sigma_\phi
  \;=\;
  \big\{\ell\,\big|\,\phi(\ell)=0\big\}.
\end{equation}
We assume that $\Sigma_\phi$ is regular in the region relevant to the integral, so that $\p_\mu\phi$ does not vanish there.

For multi-loop integrals such as Eq.~\eqref{eq:exampleform}, $\ell$ denotes the ``measured'' loop momentum, i.e., the one on which $\phi$ depends, while $f$ generally depends on all loop momenta. Derivatives with respect to unmeasured loop momenta do not act on $\delta(\phi)$ and hence give ordinary IBP identities. If $\phi$ depends on multiple loop momenta, $\ell$ stands for all of them and all contractions include a sum over them.

For a generic IBP vector $v^\mu(\ell)$, the usual total-derivative argument in dimensional regularization gives
\begin{equation}
\label{eq:ibpstart}
  0 \;=\; \int \dd^d \ell \;
  \p_\mu \!\left\{v^\mu(\ell)\,f(\ell)\,\delta(\phi(\ell))\right\},
  \qquad
  \p_\mu \equiv \frac{\p}{\p \ell^\mu}.
\end{equation}
From now on, we suppress the dependence on the loop momenta. Applying the product rule to the derivative in Eq.~\eqref{eq:ibpstart}, we obtain
\begin{equation}
\label{eq:derivative}
  0
  \;=\;
  \int \dd^d \ell \;
  \delta(\phi)\,\p_\mu\!\left(v^\mu f\right)
  +
  \int \dd^d \ell \;
  f\,v^\mu\p_\mu\delta(\phi),
\end{equation}
where
\begin{equation}
\label{eq:gradientdelta}
  \p_\mu\delta(\phi)
  \;=\;
  \delta'(\phi)\,\p_\mu\phi.
\end{equation}
Thus, for a generic $v^\mu$, the total derivative generates a derivative of
the measurement distribution. The latter can be avoided by choosing the IBP
vector tangent to the measurement hypersurface.

We first consider vector fields $t^\mu(\ell)$ satisfying the exact tangency
condition
\begin{equation}
\label{eq:tangentcondition}
  t^\mu\p_\mu\phi=0.
\end{equation}
A weaker tangency condition will be discussed in
Section~\ref{sec:weakprojector}. Applying the total-derivative argument
directly with $t^\mu$ gives
\begin{align}
  0
  &=
  \int \dd^d\ell\,
  \p_\mu\!\left\{t^\mu f\,\delta(\phi)\right\}
  \nonumber\\
  &=
  \int \dd^d\ell\,
  \delta(\phi)\,\p_\mu\!\left(t^\mu f\right)
  +
  \int \dd^d\ell\,
  f\,\delta'(\phi)\,
  t^\mu\p_\mu\phi.
\end{align}
The second term vanishes by Eq.~\eqref{eq:tangentcondition}, and one obtains
\begin{equation}
\label{eq:IBPproj}
  0
  =
  \int \dd^d\ell\,
  \delta(\phi)\,
  \p_\mu\!\left(t^\mu f\right).
\end{equation}
Thus, any vector field satisfying the exact tangency condition
\eqref{eq:tangentcondition} generates an IBP identity of the usual form, with
the measurement delta function carried along as an overall factor and never
differentiated. IBP identities of this type were previously constructed for
the special case of angular integrals in
Refs.~\cite{HaugWunder2024,HaugSmirnovWunder2025}, where a measure of the form
$\dd^{d-1}k\,\delta(\vec k^{\,2}-1)$ confines the integration to the unit
sphere and Eq.~\eqref{eq:tangentcondition} reduces to
$\vec t\cdot\vec k=0$.

To generate such tangent vectors systematically, introduce a vector field
$W^\mu(\ell)$ satisfying
\begin{equation}
\label{eq:Wdef}
  W^\mu(\ell)\,\p_\mu \phi(\ell) \;=\; 1.
\end{equation}
For any candidate IBP vector $v^\mu$, the combination
\begin{equation}
\label{eq:tfromW}
  t^\mu
  \;\equiv\;
  v^\mu-W^\mu(v\!\cdot\!\p\phi)
\end{equation}
satisfies Eq.~\eqref{eq:tangentcondition} by construction and may therefore be inserted directly into Eq.~\eqref{eq:IBPproj}. A canonical choice is
\begin{equation}
\label{eq:canonical}
  W^\mu=\frac{\p^\mu\phi}{(\p\phi)^2},
\end{equation}
provided $(\p\phi)^2\neq0$ on the support of $\delta(\phi)$. The inverse normal $W^\mu$ is not unique, and different choices generate different tangent vectors through Eq.~\eqref{eq:tfromW}. In practice, this freedom can be used to adapt the IBP vectors to the propagator structure of a given integral family.

\subsection{Transverse-momentum measurement}
\label{sec:pT}
We now specialize the construction to the transverse-momentum measurement, with constraint
\begin{equation}
\label{eq:pTdelta}
  \phi(\ell) \;=\; \frac{4}{s}(p_1\!\cdot\!\ell)(p_2\!\cdot\!\ell) - (p_T^2+m^2),
\end{equation}
where $\ell$ is the momentum of a particle of mass $m$, and $p_1,p_2$ are incoming momenta defining the beam directions, with $p_1^2=p_2^2=0$ and $s=2\,p_1\!\cdot\!p_2$. Its gradient is a linear combination of $p_1$ and $p_2$ alone,
\begin{equation}
      \p_\mu\phi(\ell) \;=\; \frac{4}{s}\Big[(p_2\!\cdot\!\ell)\,p_{1\mu} + (p_1\!\cdot\!\ell)\,p_{2\mu}\Big].
\end{equation}
The canonical choice for the inverse normal is
\begin{equation}
\label{eq:WpT}
  W^\mu(\ell)
  \;=\;
  \frac{1}{4}
  \left(
    \frac{p_1^\mu}{p_1\!\cdot\!\ell}
    +
    \frac{p_2^\mu}{p_2\!\cdot\!\ell}
  \right),
\end{equation}
which satisfies $W\!\cdot\!\p\phi=1$. Using Eq.~\eqref{eq:tfromW}, each standard generator $v\in\{\ell_i,p_1,p_2\}$ gives a tangent vector
\begin{equation}
\label{eq:tpT}
  t^\mu
  =
  v^\mu-W^\mu(v\!\cdot\!\p\phi)
  =
  v^\mu-A_1\,p_1^\mu-A_2\,p_2^\mu,
\end{equation}
with
\begin{equation}
\label{eq:A12}
 A_1=\frac{1}{s}\left[\frac{(p_2\!\cdot\!\ell)(p_1\!\cdot\!v)}{p_1\!\cdot\!\ell}+p_2\!\cdot\!v\right],
 \qquad
 A_2=\frac{1}{s}\left[\frac{(p_1\!\cdot\!\ell)(p_2\!\cdot\!v)}{p_2\!\cdot\!\ell}+p_1\!\cdot\!v\right].
\end{equation}

A practical complication is that the choice \eqref{eq:WpT} introduces inverse powers of both beam projections. If either of those scalar products is reducible, it is represented by a linear combination of propagators and kinematic invariants. Placing the full reducible expression in a denominator would then introduce the inverse of such a linear combination, which cannot be encoded through shifts in propagator indices and therefore obstructs the reduction. The freedom in choosing $W^\mu$, and hence in choosing the tangent vector generated by Eq.~\eqref{eq:tfromW}, provides several alternatives. For example, the one-sided inverse normals
\begin{equation}
\label{eq:Wonesided}
  W_1^\mu
  \;=\;
  \frac{p_1^\mu}{2\,p_1\!\cdot\!\ell},
  \qquad
  W_2^\mu
  \;=\;
  \frac{p_2^\mu}{2\,p_2\!\cdot\!\ell}
\end{equation}
also satisfy $W_{1,2}\!\cdot\!\p\phi=1$ and are preferable when the scalar product in the denominator is irreducible.

Similarly, one can take advantage of the fact that on the support of the delta function,
\begin{equation}
\label{eq:algebraic}
  (p_1\!\cdot\!\ell)
  (p_2\!\cdot\!\ell)
  \;=\;
  \frac{s}{4}(p_T^2+m^2),
\end{equation}
so that an inverse power of one reducible scalar product may be traded for a positive power of the other,
\begin{equation}
  \frac{1}{p_2\!\cdot\!\ell}
  \;=\;
  \frac{4(p_1\!\cdot\!\ell)}
       {s\,(p_T^2+m^2)},
\end{equation}
and analogously under $p_1\leftrightarrow p_2$. However, this substitution must not be made when constructing $t^\mu$ before the derivative in Eq.~\eqref{eq:IBPproj} is taken. Replacing $W^\mu$ by an expression that agrees with it only on the measurement surface $\phi=0$ generally produces a vector that is tangent only on the support of the delta function. The difference proportional to $\phi$ can generate a surviving contribution after differentiation which would otherwise be dropped. Hence, Eq.~\eqref{eq:algebraic} should be imposed only after the differentiation has been carried out, i.e., as an additional algebraic relation in the IBP system.

It is nonetheless possible to construct a modified tangent vector that incorporates the measurement constraint without imposing it prematurely and is polynomial in the relevant scalar products, i.e., it has no scalar products in the denominator. This will be discussed in the following subsection.

\subsection{Polynomial formulation}
\label{sec:weakprojector}

The exact tangency condition $t\!\cdot\!\p\phi=0$ is stronger than necessary for constructing an IBP identity in the presence of $\delta(\phi)$. It is sufficient for the vector to be tangent modulo the measurement constraint. To see this, consider a vector field $\widetilde t^\mu$ satisfying
\begin{equation}
\label{eq:weaktangent}
  \widetilde t^\mu\p_\mu\phi
  \;=\;
  h\,\phi,
\end{equation}
with $h$ a regular function of $\ell$. Using the distributional identity $x\,\delta'(x)=-\delta(x)$, we have
\begin{equation}
  \widetilde t^\mu\p_\mu\delta(\phi)
  =
  h\,\phi\,\delta'(\phi)
  =
  -h\,\delta(\phi),
\end{equation}
and applying the total-derivative argument directly to $\widetilde t^\mu$ gives
\begin{equation}
\label{eq:weakIBPt}
  0
  \;=\;
  \int\dd^d\ell\;\delta(\phi)
  \left[
    \p_\mu\!\left(f\widetilde t^\mu\right)
    -h f
  \right].
\end{equation}
Thus, a vector that is tangent only on the support of the delta function still generates a closed IBP relation, provided the off-surface deviation in Eq.~\eqref{eq:weaktangent} is kept.

Such vectors can again be generated from the usual IBP vectors. Let $\widetilde W^\mu$ satisfy the weaker inverse-normal condition
\begin{equation}
\label{eq:weakWgeneral}
  \widetilde W^\mu\p_\mu\phi
  \;=\;
  1+g\phi,
\end{equation}
With $g$ a regular function of $\ell$. Defining
\begin{equation}
\label{eq:weaktfromW}
  \widetilde t^\mu
  \;\equiv\;
  v^\mu-\widetilde W^\mu(v\!\cdot\!\p\phi),
\end{equation}
one finds
\begin{equation}
  \widetilde t^\mu\p_\mu\phi
  =
  -g\phi\,(v\!\cdot\!\p\phi).
\end{equation}
Equation~\eqref{eq:weakIBPt} therefore becomes
\begin{equation}
\label{eq:weakIBPgeneral}
  0
  \;=\;
  \int\dd^d\ell\;\delta(\phi)
  \left\{
    \p_\mu\!\left(f\widetilde t^\mu\right)
    +g\,f\,(v\!\cdot\!\p\phi)
  \right\}.
\end{equation}
This is the weakly tangent analogue of Eq.~\eqref{eq:IBPproj}.

A useful simplification occurs when the weak inverse normal additionally satisfies
\begin{equation}
\label{eq:divmatch}
  \p_\mu\widetilde W^\mu=g.
\end{equation}
Indeed, using Eq.~\eqref{eq:weaktfromW}, the integrand of
Eq.~\eqref{eq:weakIBPgeneral} can then be written as
\begin{equation}
\label{eq:weakIBPsimple}
  \p_\mu(fv^\mu)
  -
  \widetilde W^\mu
  \p_\mu\!\left[f(v\!\cdot\!\p\phi)\right].
\end{equation}
Condition \eqref{eq:divmatch} is not required for the weak construction, but selects particularly simple weak inverse normals when such a choice is available.

For the $p_T$ constraint \eqref{eq:pTdelta}, the measurement gradient obeys
\begin{equation}
\label{eq:gradientnormpT}
  (\p\phi)^2
  \;=\;
  \frac{16}{s}
  (p_1\!\cdot\!\ell)
  (p_2\!\cdot\!\ell)
  \;=\;
  4\left[\phi+(p_T^2+m^2)\right].
\end{equation}
This immediately suggests the weak inverse normal
\begin{equation}
\label{eq:Wpolynomial}
  \widetilde W^\mu
  \;=\;
  \frac{\p^\mu\phi}{4(p_T^2+m^2)}
  \;=\;
  \frac{
    (p_2\!\cdot\!\ell)p_1^\mu
    +
    (p_1\!\cdot\!\ell)p_2^\mu
  }{
    s\,(p_T^2+m^2)
  },
\end{equation}
which contains no inverse powers of the beam scalar products and satisfies
\begin{equation}
\label{eq:Wpolynomialproperties}
  \widetilde W^\mu\p_\mu\phi
  =
  1+\frac{\phi}{p_T^2+m^2},
  \qquad
  \p_\mu\widetilde W^\mu
  =
  \frac{1}{p_T^2+m^2}.
\end{equation}
Thus, $g=1/(p_T^2+m^2)$ and condition \eqref{eq:divmatch} are satisfied. The corresponding vector $\widetilde t^\mu$ obeys
\begin{equation}
  \widetilde t\!\cdot\!\p\phi
  =
  -\frac{\phi}{p_T^2+m^2}(v\!\cdot\!\p\phi),
\end{equation}
and Eq.~\eqref{eq:weakIBPsimple} therefore applies directly. All dependence on the loop momentum is polynomial, with the only additional denominator given by the kinematic factor $1/[s(p_T^2+m^2)]$.

The correction in Eq.~\eqref{eq:weakIBPgeneral} is precisely the term that would be missed by imposing the support relation \eqref{eq:algebraic} inside an exact tangent vector before differentiation. The weak construction therefore provides a polynomial alternative to the exact tangent-vector formulation of Section~\ref{sec:pT} while retaining the usual set of IBP vectors.

The construction is not specific to the single-particle transverse-momentum constraint considered here. Other non-linear observables, including the transverse recoil of a multiparticle system, rapidity differences, and energy-energy-correlator measurements, admit similarly simple tangent constructions from exact or weak inverse normals. Explicit examples are collected in Appendix~\ref{app:othermeasurements}.

\subsection{Differential equations with measurement constraints}
\label{sec:DE}
Differential equations for master integrals follow the standard strategy: differentiate with respect to a kinematic variable $x$ and reduce the resulting integrals back to the master basis. For measurement-constrained integrals of the form \eqref{eq:Imeas}, this implies
\begin{align}
\label{eq:dIdx_raw}
  \frac{\dd}{\dd x} I[f,\delta\left(\phi\right)]
  &\;=\; \int \dd^d\ell\;\frac{\dd}{\dd x}\!\left(f\,\delta(\phi)\right) \nonumber\\
  &\;=\; \int \dd^d\ell\;\delta(\phi)\,\frac{\dd f}{\dd x}
     + \int \dd^d\ell\;f\,\delta'(\phi)\,\frac{\dd \phi}{\dd x}.
\end{align}
Using any inverse normal $W^\mu$ satisfying Eq.~\eqref{eq:Wdef}, so that $\delta'(\phi)=W^\mu\p_\mu \delta(\phi)$, and integrating by parts yields
\begin{equation}
\label{eq:dIdx_final}
  \frac{\dd}{\dd x} I[f,\delta\left(\phi\right)]
  \;=\;
  \int \dd^d\ell\;\delta(\phi)\left[
    \frac{\dd f}{\dd x}
    - \p_\mu\!\left(f\,W^\mu\,\frac{\dd \phi}{\dd x}\right)
  \right].
\end{equation}
The right-hand side involves only ordinary integrands multiplied by $\delta(\phi)$ and can be reduced using the tangent-vector IBP identities \eqref{eq:IBPproj}.

For $x=p_T^2$, the integrand $f$ has no explicit $p_T^2$ dependence and $\p\phi/\p p_T^2=-1$. The weak inverse normal \eqref{eq:Wpolynomial} may therefore be used directly. Since it satisfies condition \eqref{eq:divmatch}, the weak-normal correction reduces the derivative to
\begin{align}
\label{eq:pT2polynomialDE}
  \frac{\p I}{\p p_T^2}
  &\;=\;\int\dd^d\ell\;\delta(\phi)\,
    \widetilde W^\mu\p_\mu f \nonumber\\
  &\;=\;\frac{1}{s(p_T^2+m^2)}
    \int\dd^d\ell\;\delta(\phi)
    \left[
      (p_2\!\cdot\!\ell)\,p_1\!\cdot\!\p_\ell
      +(p_1\!\cdot\!\ell)\,p_2\!\cdot\!\p_\ell
    \right]f.
\end{align}
Equation \eqref{eq:pT2polynomialDE} contains no inverse beam scalar products and provides a direct polynomial differential operator.

For $x=s$, the second term in Eq.~\eqref{eq:dIdx_final} vanishes. Hence,
\begin{equation}
\label{eq:shatDE}
  \frac{\p I}{\p s}
  =
  \frac{1}{2s}
  \int\dd^d\ell\;\delta(\phi)
  \left(
    p_1^\mu\frac{\p}{\p p_1^\mu}
    +
    p_2^\mu\frac{\p}{\p p_2^\mu}
  \right)f .
\end{equation}
The resulting integrals are reduced with the same tangent-vector IBP identities
\eqref{eq:IBPproj}.

Differential equations obtained in this way must be supplemented by boundary values. The explicit construction is described in Section~\ref{sec:unfoldingBC}.

\section{Implementation and validation}
\label{sec:applications}
The method was applied to two calculations: the Higgs transverse-momentum distribution in $pp\to t\bar tH$ at leading order and the single-inclusive top-quark transverse-momentum distribution in $pp\to t\bar t$ at next-to-leading order in QCD. Both are inclusive over all other final-state particles and defined without kinematic cuts or jet clustering. Top quarks and the Higgs boson are treated as stable on-shell particles, so that both distributions are infrared safe and are described by the single-particle measurement function~\eqref{eq:pTdelta}, with $\ell$ the momentum of the observed particle. For $t\bar t$ production we consider the charge-averaged distribution, $\tfrac12\left(\dd\sigma/\dd p_{T,t}\,+\,\dd\sigma/\dd p_{T,\bar t}\right)$, as in Ref.~\cite{Czakon:2016ptspectrum}. Diagrams for the squared amplitudes are generated with \textsc{Qgraf}~\cite{Nogueira1993} and then mapped onto integral families in \textsc{Form}~\cite{Vermaseren2000,DaviesEtAl2026FORM5}. The delta function is not introduced at the amplitude level, but only enters through the modified IBP generation, the differential operators of Section~\ref{sec:DE}, and the boundary condition construction described below.

We incorporated the tangent-vector IBP identities directly into \textsc{Kira}.
For the loop momentum entering the measurement function, each standard generator $v\in\{\ell_i,p_j\}$ is replaced by a tangent vector constructed according to Eq.~\eqref{eq:tfromW}. The resulting identities are of the form Eq.~\eqref{eq:IBPproj}. Lorentz-invariance identities are unchanged, since for two external momenta they are generated by the boost along the beam axis, under which $\phi$ is invariant.

The measurement constraint also restricts the symmetry relations that can be used in the Laporta reduction. A transformation $\mathcal T$ that preserves the propagator family and the ordinary cut conditions is a symmetry of an integral carrying $\delta(\phi)$ only if it also preserves the measurement constraint. If more generally
\begin{equation}
\label{eq:measurement-symmetry}
  \phi(\mathcal T\ell,\mathcal T p)
  =
  c\,\phi(\ell,p),
\end{equation}
with non-zero constant $c$, the delta distribution acquires a Jacobian
$|c|^{-1}$. For the
fixed-$p_T$ constraint, exchanging the incoming momenta $p_1\leftrightarrow p_2$ is compatible with the measurement, whereas a transformation exchanging cut momenta need not be, even if it preserves the propagator family in the absence of the measurement constraint. Surviving cross-family relations can be used in the reduction in the usual way.

The additional scalar-product factors introduced when constructing the tangent vectors can increase the index shifts in the IBP identities and hence the seed depth required for the Laporta reduction. The reductions are performed once for each contribution using finite-field reconstruction~\cite{KlappertLange2020,KlappertKleinLange2021}, and the sizes of the resulting master-integral bases are summarized in Table~\ref{tab:master-systems}.

\begin{table}[t]
\centering
\begin{tabular}{lc}
\toprule
contribution & number of master integrals \\
\midrule
$t\bar tH$ LO, $q\bar q$         & 13 \\
$t\bar tH$ LO, $gg$              & 26 \\
$t\bar t$ NLO, $q\bar q$ real    & 35 \\
$t\bar t$ NLO, $q\bar q$ virtual & 16 \\
$t\bar t$ NLO, $gg$ real         & 269 \\
$t\bar t$ NLO, $gg$ virtual      & 43 \\
$t\bar t$ NLO, $qg/\bar qg$ real & 31 \\
\bottomrule
\end{tabular}
\caption{Numbers of master integrals appearing in the reductions used for the application to $pp\to t\bar tH$ and $pp\to t\bar t$.}
\label{tab:master-systems}
\end{table}

\subsection{Boundary conditions for differential equations}
\label{sec:unfoldingBC}

Boundary conditions can in principle be obtained by a variety of methods and are not tied to the construction described above. For the applications considered here, a straightforward way is to partially undo the reverse-unitarity representation. The cut associated with the on-shell particle is restored to
its on-shell phase-space form and integrated together with the measurement delta function. We refer to this as ``unfolding'' the observed-particle cut. For the two-loop integrals considered here, this lowers the loop
order and leaves only the phase-space variables not fixed by the on-shell and measurement constraints.

When one remaining longitudinal degree of freedom is parameterized by a
rapidity $y$, the resulting boundary integrals have the schematic form
\begin{equation}
\label{eq:unfoldform}
  I(s,p_T^2,\varepsilon)
  =
  \mathcal N(s,p_T^2,\varepsilon)
  \int^{y_+}_{y_-} \dd y\,
  \sum_i
  r_{i}(y,s,p_T^2,\varepsilon)\,
  J_i(y,s,p_T^2,\varepsilon).
\end{equation}
Here $\mathcal N$ collects the phase-space normalization, the
$r_{i}$ are rational functions of the kinematics and $\varepsilon$ generated
by the unfolding and reduction, and the $J_i$ are ordinary cut loop integrals of one fewer loop. In the real-emission
$t\bar t$ contribution, unfolding the observed top cut leaves a continuous
rapidity integral of the form \eqref{eq:unfoldform}. For the virtual
$t\bar t$ contribution, the two massive cuts together with the $p_T$-constraint localize the top momentum to two longitudinal branches, so
that no continuous rapidity integration remains. In $t\bar tH$ production the
Higgs cut is unfolded instead, leaving one-loop cut integrals for the
$t\bar t$ system.

The rapidity integrals can contain dimensionally regulated endpoint
singularities, for which the Laurent expansion in $\varepsilon$ does not in
general commute with the remaining integration. We therefore subtract the
endpoint behavior at fixed $\varepsilon$, integrate the finite remainder
numerically, and add the subtracted terms back analytically. Details are given in Appendix~\ref{app:boundarydetails}. The unfolding can
be evaluated at generic points in the physical region, which allows the systems of differential equations to be initialized away from singular kinematic limits. The boundary values can subsequently be evolved in $(s,p_T^2)$ using the differential operators of Section~\ref{sec:DE}.

\subsection{Evolution and hadronic cross sections}
\label{sec:evolution}

The differential equations are solved numerically. For the NLO $t\bar t$ contributions, the differential-equation matrices are rational functions of $s$, $p_T^2$, and $\varepsilon$.
Where necessary, we change the basis of master integrals, by rescaling with powers of $\varepsilon$, such that the resulting $\varepsilon$-expanded systems are free of poles in $\varepsilon$. The equations for the Laurent coefficients are then integrated with an adaptive Runge--Kutta--Fehlberg method of order 7(8) using 50-digit arithmetic. The $t\bar tH$ integrals are finite and are evolved at $d=4$. Each contribution is evolved from a single boundary point.
For the $t\bar t$ contributions, direct evaluations at selected points away from the boundary point agree with the evolved results to $2\times10^{-5}$ or better.

In $t\bar t$ production, the UV-renormalized virtual, real-emission, and mass-factorization contributions are added at fixed $(s,p_T^2)$ in dimensional regularization, and their poles in $\varepsilon$ cancel in the sum (see Section~\ref{sec:validation}). For the hadronic spectra, the partonic cross sections are evaluated on grids in $(s,p_T)$ and interpolated in $s$ before convolution with the PDF. The interpolation and the treatment of the regions outside the numerical grids were checked against the NLO single-particle-inclusive kernels of Ref.~\cite{NasonDawsonEllis1989} and found to have a negligible effect on the resulting distributions.

\subsection{Validation}
\label{sec:validation}

\begin{figure}[t]
\centering
\begin{subfigure}[t]{0.48\textwidth}
\centering
\includegraphics{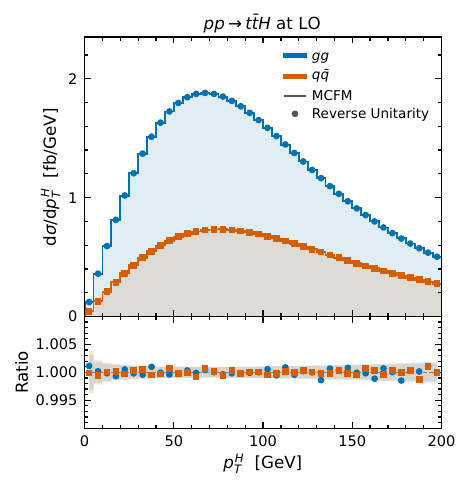}
\caption{}
\label{fig:val-tth}
\end{subfigure}
\hfill
\begin{subfigure}[t]{0.48\textwidth}
\centering
\includegraphics{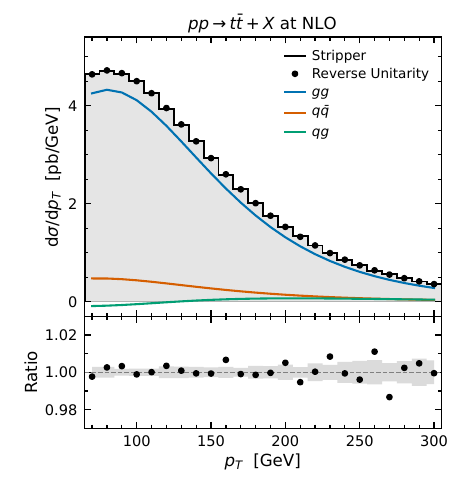}
\caption{}
\label{fig:val-ttbar}
\end{subfigure}
\caption{Validation against independent calculations at $\sqrt{S}=13$~TeV using the central member of the $\texttt{NNPDF30\_nlo\_as\_0118}$ set. 
(a) LO Higgs transverse-momentum distribution in $pp\to t\bar tH$ in the $gg$ and $q\bar q$ channels, compared with \textsc{MCFM} \cite{CampbellNeumann2019}. The lower panel shows the ratio of the bin-averaged result to \textsc{MCFM}; the bands indicate its Monte Carlo statistical uncertainty. 
(b) NLO top-quark transverse-momentum distribution in $pp\to t\bar t+X$, averaged over top and antitop, compared with \textsc{Stripper} \cite{Czakon2010,CzakonHeymes2014}. The curves show the $gg$, $q\bar q$, and $qg$ contributions, and the lower panel shows the ratio of the bin-averaged result to \textsc{Stripper}, with the band indicating its statistical uncertainty.}
\label{fig:validation}
\end{figure}

For the hadronic comparisons we use $\sqrt{S}=13~\mathrm{TeV}$ and the central member of the \texttt{NNPDF30\_nlo\_as\_0118} parton distribution set \cite{NNPDF2015}. For $t\bar t$ production we set $m_t=173~\mathrm{GeV}$ and $\mu_R=\mu_F=m_t$, while for $t\bar tH$ production we additionally take $m_H=125~\mathrm{GeV}$ and $\mu_R=\mu_F=(2m_t+m_H)/2$. The same input parameters, scales, and parton distributions are used in the corresponding reference calculations.

In both cases the complete hadronic calculation, including the convolution with parton distributions, was compared with an independent implementation. For $t\bar tH$, Fig.~\ref{fig:val-tth} shows the ratio of our results, averaged over each bin, to \textsc{MCFM}~\cite{CampbellNeumann2019}. In both channels the ratios lie between 0.9985 and 1.0011, consistent with the statistical uncertainty of \textsc{MCFM} of 0.1--0.4\% per bin. Integrated over $0<p_T<200$~GeV, the two calculations agree within the statistical uncertainty of \textsc{MCFM} of 0.02\%.

The NLO $t\bar t$ calculation provides a more stringent validation of the framework. The real and virtual contributions are reduced and evaluated independently, and we compare their poles in $\varepsilon$ at fixed $(s,p_T^2)$ for a set of reference points in each channel. In the $q\bar q$ and $gg$ channels, in units of $(\alpha_s/2\pi)\,\sigma_B$, with $\sigma_B$ the Born cross section at the same kinematic point, the coefficients of $1/\varepsilon^2$ in the virtual contributions are expected to be $-2C_F$ and $-2C_A$, respectively, with the opposite coefficients in the real contributions. The computed coefficients agree with these values to better than $10^{-13}$. The single poles cancel in the sum of the real, virtual, and mass-factorization contributions to better than $10^{-7}$ relative to the individual terms. The $qg$ channel has no virtual correction and no soft singularity at this order. Its real contribution has no double pole, while its single pole is purely of initial-state collinear origin and cancels against the mass-factorization counterterm to the same accuracy.

The finite partonic coefficients were independently compared with the one-particle-inclusive heavy-quark kernels of Ref.~\cite{NasonDawsonEllis1989}, as implemented in \textsc{FONLL} \cite{CacciariGrecoNason1998}. The relative agreement is better than $10^{-9}$ in the $q\bar q$ channel and $10^{-4}$ in the $gg$ and $qg$ channels. Finally, the full hadronic NLO spectrum was compared with an independent calculation using the \textsc{Stripper} framework \cite{Czakon2010,CzakonHeymes2014}, shown in Fig.~\ref{fig:val-ttbar}, with per-bin agreement consistent with the Monte Carlo uncertainties in all channels.

Together, these comparisons test the complete calculation, from the tangent-vector IBP reduction and boundary conditions through the differential evolution to the hadronic cross section.

\subsection{Computational cost and higher orders}
\label{sec:cost}

The largest reduction, for the $gg$ real contribution to $t\bar t$ production, took about 30 hours on a workstation, using up to 48 cores and a peak memory of 213~GB. Each of the other NLO reductions took at most eight hours. The boundary values are inexpensive in comparison. The evolution of the 1728 values of $(s,p_T)$ entering the NLO $t\bar t$ spectrum took about 200 CPU-hours, and the \textsc{Stripper} calculation for Fig.~\ref{fig:val-ttbar} took about 50 CPU-hours on the same workstation, for a statistical uncertainty of about $0.3\%$ per bin. The reductions are one-time costs for a given process and observable, whereas the evolution is repeated for each kinematic point. Since the partonic cross sections are independent of the PDF, a change of PDFs at fixed scales requires only a new convolution, which takes minutes.

At higher orders, the cost is dominated by the IBP reduction of three-loop cut integrals, which arise at NNLO for $t\bar t$ and at NLO for $t\bar tH$ from all contributions whose phase space is not fixed by the cuts and the measurement. For these, the tangent vectors can increase the required seed depth, and the measurement reduces the symmetries between integral families compared with an inclusive calculation.

We considered three-loop integral families occurring in the NLO correction to $q\bar q\to t\bar tH$, with a transverse-momentum measurement on the Higgs momentum. The master integrals could be identified within a few CPU-hours. However, a complete reduction requires the reconstruction of about $10^6$ rational coefficients and could not be completed with the available IBP reducers.

The boundary values pose a second difficulty, since the method of Section~\ref{sec:unfoldingBC} would leave two-loop integrals under the remaining phase-space integration. For these contributions, a strategy such as the auxiliary-mass-flow method~\cite{LiuMaWang2018,LiuMa2023} in its formulation for phase-space integrals~\cite{LiuMaTaoZhang2021} appears most promising. The measurement function does not depend on the auxiliary mass and the resulting differential equations could therefore be reduced with the tangent-vector IBP identities. We leave this for future work.

\section{Conclusions}
\label{sec:conclusions}

We have derived IBP identities for dimensionally regulated
integrals carrying a measurement delta function $\delta(\phi(\ell))$, with
$\phi$ a general non-linear function of the loop momenta. The construction rests on the observation that the derivative of the measurement delta function drops out of the total-derivative identity when the IBP vector is tangent to the hypersurface $\phi=0$.
Choosing the IBP vector to be tangent to the measurement hypersurface therefore yields a closed system in which the measurement delta is carried along as an overall factor and is never differentiated. Hence, the resulting relations can be passed to a conventional Laporta solver once they have been generated.

We implemented the method for the transverse-momentum measurement constraint and combined the resulting reductions with boundary values obtained by unfolding the observed-particle cut into lower-loop integrals. The master integrals can then be evolved from generic points in the physical region through differential equations in $(s,p_T^2)$.

The complete framework was applied to and validated for the Higgs transverse-momentum spectrum in $pp\to t\bar tH$ at leading order and to the single-inclusive top-quark transverse-momentum spectrum in $pp\to t\bar t$ at next-to-leading order in QCD. 
At higher orders, the main obstacles are the reduction of three-loop cut integrals and the determination of their boundary values.
The construction is not restricted to the single-particle transverse momentum considered in the applications. As illustrated in Appendix~\ref{app:othermeasurements}, exact or weakly tangent vectors can also be constructed for other non-linear constraints, including multiparticle transverse recoil, rapidity differences and energy-energy correlators.

\acknowledgments
I thank Micha\l{} Czakon for valuable discussions and comments on the manuscript.
This research was supported by the Deutsche Forschungsgemeinschaft (DFG) under grant 400140256 -- GRK 2497: \textit{The Physics of the Heaviest Particles at the LHC}.

\appendix
\section{Details of the unfolded boundary integrals}
\label{app:boundarydetails}

\subsection{Real-emission unfolding and normalization}

For a momentum $\ell$ with
\begin{equation}
  \phi(\ell)
  =
  \frac{4}{s}
  (p_1\!\cdot\!\ell)(p_2\!\cdot\!\ell)
  -(p_T^2+m^2),
\end{equation}
the on-shell condition and the measurement delta fix the transverse momentum
while leaving the rapidity continuous. After the transverse angular integration,
\begin{equation}
\label{eq:unfoldmeasure}
  \int\dd^d\ell\,
  \delta_+(\ell^2-m^2)\,
  \delta\bigl(\phi(\ell)\bigr)\,
  h(\ell)
  =
  \frac{\pi^{d/2-1}}
       {2\,\Gamma(d/2-1)}
  (p_T^2)^{d/2-2}
  \int\dd y\;h\bigl(p(y)\bigr),
\end{equation}
where the remaining cut conditions determine the rapidity range. For the real-emission
$t\bar t$ kinematics, we have
\begin{equation}
\label{eq:ymax}
  y_{\max}
  =
  \operatorname{arccosh}
  \left(
    \frac{\sqrt{s}}{2m_T}
  \right),
  \qquad
  m_T^2=m_t^2+p_T^2.
\end{equation}

Including the reverse-unitarity normalization of the unfolded cut, the universal
factor multiplying the rapidity integral is
\begin{align}
\label{eq:unfoldingprefactor}
  \mathcal N_y
  &=
  -\frac{i}{4(2\pi)^{d-1}}
  \frac{2\pi^{d/2-1}}{\Gamma(d/2-1)}
  (p_T^2)^{d/2-2}
  \nonumber\\
  &=
  -\frac{i}{16\pi^2}
  \frac{(4\pi/p_T^2)^\varepsilon}{\Gamma(1-\varepsilon)}.
\end{align}
Scheme-dependent loop-normalization factors are applied together with the
corresponding amplitude conventions and are not included in
\eqref{eq:unfoldingprefactor}.

After unfolding, propagators that depend only on momentum $\ell$ become
ordinary functions of $y$. The remaining loop-dependent denominators are ordinary propagators forming one-loop cut families.

\subsection{Endpoint subtraction before the
\texorpdfstring{$\varepsilon$}{epsilon} expansion}

For the unfolded representation \eqref{eq:unfoldform}, it is useful to denote the
rapidity integrand by
\begin{equation}
  \mathcal I(y,s,p_T^2,\varepsilon)
  \equiv
  \sum_i
  r_i(y,s,p_T^2,\varepsilon)\,
  J_i(y,s,p_T^2,\varepsilon).
\end{equation}
At $y\to\pm y_{\max}$ the recoiling system approaches the boundary of phase
space. For the real-emission $t\bar t$ kinematics considered here, this
corresponds to the recoiling $\bar t+g$ system reaching threshold, with the emitted gluon becoming soft. Suppressing the dependence on $s$ and $p_T^2$ in the
following, we introduce the distances from the two endpoints,
\begin{equation}
  \Delta_\pm = y_{\max}\mp y.
\end{equation}
Both the lower-loop integrals $J_i$ and their rational coefficients $r_i$ can
contribute powers of $\Delta_\pm$. It is therefore the endpoint behavior of the
complete products $r_iJ_i$, rather than that of the lower-loop integrals alone,
that determines the required subtraction. The resulting expansions at the two
endpoints have the generic form
\begin{equation}
\label{eq:endpointtower}
  \mathcal I(\Delta_\pm,\varepsilon)
  =
  \sum_{n=0}^{N}
  c_n^\pm(\varepsilon)\,
  \Delta_\pm^{p+n-2\varepsilon}
  +
  \mathcal O\!\left(
    \Delta_\pm^{p+N+1-2\varepsilon}
  \right),
\end{equation}
where the coefficients $c_n^\pm$ may differ between the two endpoints. For the
real-emission $t\bar t$ boundary integrals encountered here, endpoint terms with leading behavior
$\Delta^{-2\varepsilon}$,
$\Delta^{-1-2\varepsilon}$, and
$\Delta^{-2-2\varepsilon}$
occur.

Whenever a term in \eqref{eq:endpointtower} has $p+n\leq-1$, the expansion in
$\varepsilon$ cannot in general be interchanged with the endpoint integration.
For example, $\Delta^{-1-2\varepsilon}$ is defined by dimensional
regularization, whereas expanding it first produces terms proportional to
$\Delta^{-1}\ln^k\Delta$ that are not separately integrable at $\Delta=0$.
Consequently, poles in $\varepsilon$ can be carried entirely by the regulated
endpoint integral, while numerical quadrature of a prematurely expanded
integrand can converge to a finite, stable, and incorrect value.

The singular part of the endpoint expansion is therefore subtracted at fixed
$\varepsilon$. The subtracted remainder can then be expanded in
$\varepsilon$ and integrated numerically, while the subtraction terms are
integrated analytically and added back. For each term in
\eqref{eq:endpointtower},
\begin{equation}
\label{eq:endpointaddback}
  \int_0^\Lambda\dd\Delta\,
  \Delta^{p+n-2\varepsilon}
  =
  \frac{\Lambda^{p+n+1-2\varepsilon}}
       {p+n+1-2\varepsilon}.
\end{equation}
For either endpoint, the corresponding distance $\Delta_\pm$ ranges from zero
to $2y_{\max}$ over the full rapidity interval, so that
$\Lambda=2y_{\max}$ in \eqref{eq:endpointaddback}.

The coefficients $c_n^\pm(\varepsilon)$ required for the subtraction terms are
obtained from the endpoint expansions of the corresponding lower-loop
integrals together with their rational prefactors. For the boundary integrals
appearing in the present calculation, these coefficients are available as
closed functions of $\varepsilon$ and are expanded to the order required by the
desired pole depth after the analytic integration.

\section{Further examples of non-linear measurement constraints}
\label{app:othermeasurements}

The construction of Section~\ref{sec:IBP} is not restricted to a single-particle transverse-momentum measurement. In this appendix we give examples of other non-linear constraints for which tangent vectors can be generated from simple exact or weak inverse normals. These examples are not pursued in this work.

\subsection{Transverse recoil of a multiparticle system}
\label{app:pairqT}

Consider a multiparticle system with total momentum
\begin{equation}
  K^\mu = \sum_i \ell_i^\mu .
\end{equation}
Its transverse recoil with respect to the two lightlike beam momenta can be fixed
by
\begin{equation}
\label{eq:pairqTconstraint}
  \phi_{q_T}(K)
  \;=\;
  \frac{4}{s}(p_1\!\cdot\!K)(p_2\!\cdot\!K)
  -\left(K^2+q_T^2\right).
\end{equation}
In contrast to the single-particle case, $K^2$ is not a fixed mass parameter
but depends on the integration momenta. Its derivative therefore contributes to the normal of the measurement surface and, as shown below, cancels the longitudinal part, leaving a purely transverse normal vector.

Introducing the decomposition
\begin{equation}
  K^\mu = K_\parallel^\mu + K_\perp^\mu,
  \qquad
  K_\parallel^\mu
  = \frac{2}{s}\left[
  \left(p_2\!\cdot\!K\right)p_1^\mu
  +
  \left(p_1\!\cdot\!K\right)p_2^\mu\right],
\end{equation}
where $K_\perp\!\cdot p_1=K_\perp\!\cdot p_2=0$, one finds
\begin{equation}
  \p_K^\mu \phi_{q_T}
  \;=\;
  -2K_\perp^\mu,
  \qquad
  (\p_K\phi_{q_T})^2
  \;=\;
  -4\bigl(\phi_{q_T}+q_T^2\bigr).
\end{equation}
Since
\begin{equation}
  K_\parallel\!\cdot\!\p_K\phi_{q_T}=0,
\end{equation}
a component proportional to $K_\parallel^\mu$ may be added to the weak inverse
normal without affecting its contraction with $\p_K\phi_{q_T}$. We use this
freedom to choose a polynomial weak inverse normal that also satisfies the
simplifying condition \eqref{eq:divmatch}. Since
\begin{equation}
  \p_K\!\cdot K_\perp=d-2,
  \qquad
  \p_K\!\cdot K_\parallel=2,
\end{equation}
the choice
\begin{equation}
\label{eq:pairqTweaknormal}
  \widetilde W_{q_T}^\mu
  =
  \frac{1}{2q_T^2}
  \left(
    K_\perp^\mu
    +\frac{4-d}{2}\,K_\parallel^\mu
  \right)
\end{equation}
satisfies
\begin{equation}
  \widetilde W_{q_T}\!\cdot\!\p_K\phi_{q_T}
  =
  1+\frac{\phi_{q_T}}{q_T^2},
  \qquad
  \p_K\!\cdot\widetilde W_{q_T}
  =
  \frac{1}{q_T^2}.
\end{equation}
Thus $g=1/q_T^2$ and condition \eqref{eq:divmatch} is fulfilled. In dimensional
regularization, $(4-d)/2=\varepsilon$, so the second term in
Eq.~\eqref{eq:pairqTweaknormal} is an $\mathcal O(\varepsilon)$ tangent contribution
chosen precisely to obtain the simplified weakly tangent identity
\eqref{eq:weakIBPsimple}.

\subsection{Rapidity differences}
\label{app:rapiditydifference}

As a second example, consider the rapidity difference between two measured
final-state momenta $\ell_1$ and $\ell_2$,
\begin{equation}
  \Delta y = y_1-y_2,
  \qquad
  y_i =
  \frac{1}{2}
  \ln\!\left(
    \frac{p_2\!\cdot\!\ell_i}{p_1\!\cdot\!\ell_i}
  \right).
\end{equation}
Fixing $\Delta y$ can be expressed through the bilinear measurement function
\begin{equation}
\label{eq:deltayconstraint}
  \phi_{\Delta y}
  =
  (p_2\!\cdot\!\ell_1)(p_1\!\cdot\!\ell_2)
  -
  \e^{2\Delta y}
  (p_1\!\cdot\!\ell_1)(p_2\!\cdot\!\ell_2).
\end{equation}
In contrast to fixing the two rapidities separately, this single constraint fixes only their difference, while the average rapidity of the two particles remains unrestricted.

Differentiating with respect to $\ell_1$ gives
\begin{equation}
\label{eq:deltaygradient}
  \p_{\ell_1}^{\mu}\phi_{\Delta y}
  =
  (p_1\!\cdot\!\ell_2)\,p_2^\mu
  -
  \e^{2\Delta y}(p_2\!\cdot\!\ell_2)\,p_1^\mu ,
\end{equation}
with
\begin{equation}
\label{eq:deltaygradientnorm}
  \left(\p_{\ell_1}\phi_{\Delta y}\right)^2
  =
  -s\,\e^{2\Delta y}
  (p_1\!\cdot\!\ell_2)(p_2\!\cdot\!\ell_2).
\end{equation}
The canonical exact inverse normal is therefore
\begin{equation}
\label{eq:deltaynormal}
  W_{\Delta y}^{\mu}
  =
  \frac{\p_{\ell_1}^{\mu}\phi_{\Delta y}}
       {\left(\p_{\ell_1}\phi_{\Delta y}\right)^2}
  =
  \frac{p_1^\mu}
       {s\,(p_1\!\cdot\!\ell_2)}
  -
  \frac{\e^{-2\Delta y}p_2^\mu}
       {s\,(p_2\!\cdot\!\ell_2)},
\end{equation}
which satisfies
\begin{equation}
  W_{\Delta y}\!\cdot\!
  \p_{\ell_1}\phi_{\Delta y}
  = 1 .
\end{equation}

\subsection{Energy-energy correlators}
\label{app:EECprojected}

Finally, consider the energy-energy-correlator measurement introduced at the beginning of Section~\ref{sec:IBP}. In the same convention, its angular constraint can be written as
\begin{equation}
\label{eq:EECconstraint_projected}
  \phi_{\rm EEC}
  \;=\;
  (1-\cos\chi)\,(p_i\!\cdot\!p)(p_j\!\cdot\!p)
  -p^2\,(p_i\!\cdot\!p_j),
\end{equation}
where $p$ is the total incoming momentum. Differentiating with respect to $p_i$
yields
\begin{equation}
  \p_{p_i}^\mu\phi_{\rm EEC}
  \;=\;
  (1-\cos\chi)(p_j\!\cdot\!p)\,p^\mu
  -p^2p_j^\mu.
\end{equation}
Contracting with $p^\mu$ gives
\begin{equation}
  p\!\cdot\!\p_{p_i}\phi_{\rm EEC}
  \;=\;
  -\cos\chi\;p^2(p_j\!\cdot\!p),
\end{equation}
so that, for $\cos\chi\neq0$,
\begin{equation}
\label{eq:EECnormal_projected}
  W_{\rm EEC}^\mu
  \;=\;
  -\frac{p^\mu}
  {\cos\chi\;p^2(p_j\!\cdot\!p)}
\end{equation}
is an exact inverse normal,
\begin{equation}
  W_{\rm EEC}\!\cdot\!\p_{p_i}\phi_{\rm EEC}
  \;=\;
  1.
\end{equation}
The singularity of the particular choice \eqref{eq:EECnormal_projected} at
$\chi=\pi/2$ is not a singularity of the measurement constraint. 
For the standard massless EEC kinematics, $p_j^2=0$. At
$\chi=\pi/2$, one therefore finds
\begin{equation}
  (\p_{p_i}\phi_{\rm EEC})^2
  =
  -p^2(p_j\!\cdot\!p)^2
  \neq 0,
\end{equation}
so that the canonical inverse normal
\begin{equation}
  W_{\rm EEC,can}^\mu
  =
  \frac{\p_{p_i}^\mu\phi_{\rm EEC}}
       {(\p_{p_i}\phi_{\rm EEC})^2}
\end{equation}
is regular at this point.

\bibliographystyle{JHEP}
\bibliography{biblio}

\end{document}